%% file: NAHeis-070103.tex
\documentclass[12pt]{article}
\usepackage{geometry}
\usepackage{amsmath,amssymb}
\usepackage[nosort]{cite}
\usepackage{float}
\usepackage{latexsym}
\usepackage{graphicx}
\usepackage{color}
\newcommand{\fft}[2]{{\frac{#1}{#2}}}

\newcommand{\Tr}{{\rm Tr\,}}

\newcommand{\beq}{\begin{equation}}
\newcommand{\be}{\begin{equation}}
\newcommand{\ee}{\end{equation}}
\newcommand{\bea}{\begin{eqnarray}}
\newcommand{\eea}{\end{eqnarray}}

\newcommand{\bb}[2]{\mbox{{\bf{#1}}}_{#2}}

\newcommand{\bIII}{\bar{\mbox{{\bf{3}}}}}
\newcommand{\bbs}[1]{\mbox{{\bf{#1}}}}
\newcommand{\bbbs}[1]{\bar{\mbox{{\bf{#1}}}}}

\newcommand{\nn}{\nonumber}

\newcommand{\zet}{{\mathbb Z}}

\makeatletter \@addtoreset{equation}{section} \makeatother

\begin{document}
\begin{titlepage}

\hbox to \hsize{\hbox{\tt hep-th/0701028}\hss
    \hbox{\small{\tt MCTP-07-01}}}
\hbox to \hsize{\hbox{ }\hss
    \hbox{\small{\tt TAUP-2844-07}}}

\vspace{1 cm}

\begin{center}

{\bf \Large Finite Heisenberg Groups from Nonabelian\\[.5cm]
Orbifold Quiver Gauge Theories}

\vspace{1 cm}
{\large $^\dagger$Benjamin A. Burrington, $^*$James T. Liu, $^*$Leopoldo A. Pando Zayas}

\vspace{.5 cm}

{\it $^\dagger$School of Physics and Astronomy,\\
The Raymond and Beverly Sackler Faculty of Exact Sciences,\\
Tel Aviv University, Ramat Aviv, 69978, Israel.}

\vspace{.5 cm}

{\it $^*$Michigan Center for Theoretical Physics,\\
Randall Laboratory of Physics, The University of Michigan\\
Ann Arbor, MI 48109--1040, USA}

\end{center}

\vspace{0.3 cm}

\begin{abstract}
A large class of orbifold quiver gauge theories admits the action of finite
Heisenberg groups of the form $\prod_i{\rm Heis}(\mathbb{Z}_{q^{\;}_i}\times
\mathbb{Z}_{q^{\;}_i})$.  For an Abelian orbifold generated by $\Gamma$,
the $\mathbb Z_{q^{\;}_i}$ shift generator in each Heisenberg group is one
cyclic factor of the Abelian group $\Gamma$.  For general non-Abelian
$\Gamma$, however, we find that the shift generators are the cyclic factors
in the Abelianization of $\Gamma$.  We explicitly show this for the case
$\Gamma=\Delta(27)$, where we construct the finite Heisenberg group
symmetries of the field theory.  These symmetries are dual to brane number
operators counting branes on homological torsion cycles, which therefore do
not commute. We compare our field theory results with string theory states
and find perfect agreement.
\end{abstract}
\end{titlepage}

\section{Introduction}

The AdS/CFT correspondence, which relates field theories with
dual string (gravitational) theories, is well on its way to
becoming firmly established as an especially important new tool for
investigating field theories outside of perturbative control.  As
a strong/weak coupling duality, AdS/CFT and its variants have opened
up a whole new window into the exploration of strongly coupled gauge
theories through their weakly coupled string counterparts.  Certainly,
much of the focus of gauge/string duality has been on the use of
appropriate string duals as a way to address important non-perturbative
issues in gauge theories.
However, given the limited technology for studying string theory in
non-trivial closed string backgrounds ({\it i.e.}~curved backgrounds
with RR flux), one may ask whether it is possible to use AdS/CFT in
the other direction to explore the structure of string theory in such
backgrounds starting from our understanding of gauge theories.

In fact, Gukov, Rangamani and Witten in~\cite{Gukov:1998kn} did just this;
based on properties of the dual orbifold quiver gauge theory, they found
the novel feature that brane number charges counting branes wrapped on
torsion cycles%
\footnote{Here we will always refer to the topological ``torsion'' groups
appearing in integer valued homology rather than notions relating to
modifications of connections on a tangent bundle.  To stress this, we will
usually refer to these topological cycles as ``torsion cycles.''}
do not commute.  In particular, they examined the duality between string
theory on AdS$_5\times S^5/\mathbb Z_3$ and the quiver gauge theory
corresponding to a stack of D3-branes placed at the singular point of
$\mathbb{R}^{1,3}\times \mathbb C^3/\mathbb Z_3$, and demonstrated that
the resulting quiver gauge theory admits a set of discrete global
symmetries that forms a Heisenberg group.  Based on this, they concluded
that, on the gravitational
side, F-string and D-string number operators do not commute when the
strings are wrapped on torsion cycles.  In fact, these operators close
on the number operator for D3-branes wrapping a torsion 3-cycle, and
it is this D3-brane number operator which plays the part of the central
extension in the Heisenberg group.

A similar observation about fluxes on torsion cycles was recently made
by Freed, Moore and Segal in \cite{Freed:2006ya,Freed:2006yc} for
generalized $\mathrm U(1)$ gauge theories.
In this case, Poincar\'e duality of the theory requires that
two gradings (electric and magnetic) of the Hilbert
space be possible.  Furthermore, in the presence of torsion cycles,
these two gradings cannot be simultaneously implemented: {\it i.e.}~electric
and magnetic charges do not commute.  As argued in \cite{Freed:2006ya},
it is precisely this effect that is responsible
for the non-commutativity of charges found in \cite{Gukov:1998kn}.

The non-commutativity in \cite{Gukov:1998kn} was found as a
non-commutativity of global symmetries of the quiver gauge
theory dual.  These discrete symmetries fall into two classes, with the
first being permutation type symmetries that map gauge groups into
gauge groups (and correspondingly bifundamentals into bifundamentals)
and the second being rephasing symmetries which act on the links in
such a manner that all trace type gauge invariant composite operators
remain unchanged under the rephasing.  We will refer to the permutation
symmetries here as $\mathcal A$ type symmetries and the rephasing
symmetries as either $\mathcal B$ (if they do not commute with
permutations) or $\mathcal C$ (if they are central) type symmetries.

In particular, for the $S^5/\mathbb Z_3$ quiver theory considered in
\cite{Gukov:1998kn}, the permutation symmetries are generated by a
`shift' operator $\mathcal A$ of order three which cyclically permutes
the three nodes of the quiver.  (Note that this operation is simply
the action of the $\mathbb Z_3$ orbifold group on the quiver itself.)
In addition, the rephasing symmetries are generated by an order
three `clock' operator $\mathcal B$.  Taken together, $\mathcal A$ and
$\mathcal B$ do not commute, and form the Heisenberg group
$\mbox{Heis}(\mathbb Z_3\times\mathbb Z_3)$ according to
\begin{equation}
\mathcal A^3=\mathcal B^3=\mathcal C^3=1,\qquad \mathcal{AB}=\mathcal{BAC},
\label{eq:ab=bac}
\end{equation}
where $\mathcal C$ is a central element associated with permutation invariant
rephasings.

The finite Heisenberg group identified above is the group of discrete
global symmetries acting on the quiver gauge theory.  However, as argued
in \cite{Gukov:1998kn}, these symmetries ought to persist in the dual
description of strings on AdS$_5\times S^5/\mathbb Z_3$.  In this dual
string picture, the $\mathcal A$ operator is mapped to the number operator
associated with wrapped F-strings, while the $\mathcal B$ and $\mathcal C$
operators correspond to number operators associated with wrapped D-strings
and D3-branes, respectively.  Although $S^5$ by itself admits no
non-trivial cycles, $S^5/\mathbb Z_3$ admits the torsion cycles
$H_1(S^5/\mathbb Z_3;\mathbb Z)=\mathbb Z_3$ and
$H_3(S^5/\mathbb Z_3;\mathbb Z)=\mathbb Z_3$, which are precisely
what are needed for strings and D3-branes to wrap.  In this case, the
non-commutativity of (\ref{eq:ab=bac}) is a consequence of the
non-commutativity of fluxes \cite{Freed:2006ya,Freed:2006yc} associated
with these cycles.  Furthermore, the fact that trace type gauge invariant
operators are inert under the $\mathcal B$ and $\mathcal C$ symmetries
indicate that their dual operators affect only states that are
non-perturbative in the $1/N$ expansion, giving additional justification
to their association with D-branes.

The work of \cite{Gukov:1998kn} was recently extended by studying other
orbifold theories \cite{Burrington:2006uu}, the inclusion of fractional
branes \cite{Burrington:2006aw}, and the effect of Seiberg duality on
these global symmetries \cite{Burrington:2006pu}.  In all these
investigations, however, the orbifolds under consideration were Abelian,
being $\mathbb Z_n$ orbifolds of either $S^5$, $T^{1,1}$ or $Y^{p,q}$.
For the case without fractional branes, the resulting discrete symmetry
groups were all found to be isomorphic to $\mbox{Heis}(\mathbb Z_n\times
\mathbb Z_n)$, which is exactly what one would expect to have been
inherited from the $\mathbb Z_n$ action of the orbifold group.

Although the bulk of the work on orbifold models focuses on the Abelian
case, in general non-Abelian orbifolds may also be constructed.  Other
than for added technical issues which arise in the latter case, there
is no fundamental distinction between Abelian and non-Abelian orbifolds.
Hence it is natural to expect that the identification of discrete global
symmetries of Abelian orbifold quiver gauge theories carries over to
non-Abelian orbifolds as well.  Demonstrating that this is in fact the
case will be the focus of the present paper.  In
particular, we examine quiver gauge theories that are dual to IIB
string theory on AdS$_5 \times S^5/\Gamma$, where now $\Gamma$ is some
non-Abelian discrete subgroup of $\mathrm{SU}(3)$, so that the gauge theory
preserves at least $\mathcal N=1$ supersymmetry.  Although we only
focus on orbifolds of the five-sphere, it is in principle possible to
use the same techniques for field theories dual to orbifolds of
other spaces which admit isometries.  One simply has to use the
regular representation of the group $\Gamma$ to act
on the gauge indices, which is the natural way of making a
``geometric'' orbifold \cite{Kachru:1998ys,Lawrence:1998ja}.
In the following, we will {\it always} be using the regular
representation of $\Gamma$ as the group acting on the gauge indices.

In the case where $\Gamma$ is Abelian, the $\mathcal{A}$ type permutation
symmetries are easy to visualize geometrically, as they simply correspond
to the action of the group $\Gamma$ mapping the image D3-branes to each other
(so that F-strings stretched between D3-branes and their images
close up to an operation of $\mathcal A$).  This immediately suggests
that the group of $\mathcal{A}$ type symmetries remains $\Gamma$,
even when extended to the non-Abelian case.  As we show below, however,
this identification is not entirely correct; instead, the actual group
generated by the $\mathcal A$ type symmetries turns out to be the
Abelianization of $\Gamma$, which we denote by $\bar\Gamma
\equiv\Gamma/[\Gamma,\Gamma]$.

Unlike the $\mathcal A$ type symmetries, which are conceptually easy
to visualize based on the action of $\Gamma$ on the quiver, the
$\mathcal B$ and $\mathcal C$ type rephasing symmetries are more
difficult to identify.  Motivated by \cite{Gukov:1998kn} as well
as the dual string picture (where $\mathcal A$ and $\mathcal B$
are related by interchanging F-strings and D-strings), it is
natural to identify the group of $\mathcal B$ type symmetries with
$\bar\Gamma$ as well.  However, the explicit identification
of the rephasing symmetries is significantly different, and rather
more intricate to work out.  Nevertheless, as demonstrated in
various examples, this general picture of $\mathcal A$ and $\mathcal B$
type operations turns out to be correct.  The $\mathcal A$ and
$\mathcal B$ symmetries do not commute, and instead close on a
set of central elements given by $\mathcal C$.  Since $\mathcal A$
and $\mathcal B$ are both identified with $\bar\Gamma$, the
group formed by these symmetries is just the finite Heisenberg group
\begin{equation}
\mathrm{Heis}(\bar\Gamma,\bar\Gamma).
\label{eq:heis}
\end{equation}
This is identified as the group of discrete global symmetries of the
$S^5/\Gamma$ orbifold quiver gauge theory.

Turning now to the dual string picture, we expect that the above
quiver gauge theory may be obtained by placing a stack of $N$
D3-branes near the orbifold singularity of $\mathbb R^{1,3}\times
\mathbb C^3/\Gamma$.  Since the near-horizon space is
AdS$_5\times S^5/\Gamma$, in this dual picture the Heisenberg group
symmetries (\ref{eq:heis}) arise from the properties of strings and
D3-branes wrapping appropriate cycles on $S^5/\Gamma$.  Since the
homology of $S^5/\Gamma$ is given by \cite{Morrison:1998cs}
\begin{equation}
H_1(S^5/\Gamma;\mathbb Z)=\bar\Gamma,\qquad
H_3(S^5/\Gamma;\mathbb Z)=\bar\Gamma,
\label{eq:homo}
\end{equation}
we indeed identify the appropriate cycles for F-strings, D-strings and
D3-branes to wrap.  In particular, since the string duals of $\mathcal A$
and $\mathcal B$ are the number operators counting wrapped F-strings and
D-strings, they both must independently form groups isomorphic to $H_1$,
which as we see is just $\bar\Gamma$.  This of course agrees with
the quiver result.

The final part of the duality picture is to identify $\mathcal C$
with wrapped D3-branes, and hence the group generated by $\mathcal C$
with $H_3$ (which again is just $\bar\Gamma$).  This now allows
us to make a stronger statement on the form of the Heisenberg group
(\ref{eq:heis}).  Since $\bar\Gamma$ is Abelian, it may be decomposed
into  a set of cyclic groups
\be
\label{Abelianization}
\bar\Gamma=\zet_{a^{\;}_1}\otimes \zet_{a^{\;}_2}\otimes \cdots,
\ee
where each factor is associated with a torsion cycle in $H_1(S^5/\Gamma)$.
Because $\mathcal A$, $\mathcal B$ and $\mathcal C$ are all identified
with $\bar\Gamma$, the central extension implicit in (\ref{eq:heis})
may be more explicitly stated in the decomposition
\begin{equation}
\mathrm{Heis}(\bar\Gamma,\bar\Gamma)=
\mathrm{Heis}(\zet_{a^{\;}_1},\zet_{a^{\;}_1})\otimes
\mathrm{Heis}(\zet_{a^{\;}_2},\zet_{a^{\;}_2})\otimes\cdots.
\label{eq:manyheis}
\end{equation}
Each individual factor in this decomposition is connected to the
non-commutativity of fluxes on that particular torsion cycle
\cite{Freed:2006ya,Freed:2006yc}.  (Since we take $\Gamma$
to be a discrete subgroup of $\mathrm{SU}(3)$, there are only a limited
number of torsion cycles that may arise in practice.)
Our main result can then be stated as:
\begin{quote}
Quiver gauge theories obtained as worldvolume theories on a stack of $N$
D3-branes placed at the singularity of $\mathbb{C}^3/\Gamma$ where $\Gamma$
is a (possibly non-Abelian) finite discrete subgroup of $\mathrm{SU}(3)$
admit the action of global symmetries generating a group of the form
(\ref{eq:manyheis}) where the factors $\zet_{a^{\;}_i}$ are given by the Abelianization of $\Gamma$ as in (\ref{Abelianization}).
\end{quote}

The structure of the paper is as follows. In section \ref{sec:nonabelian}
we review the construction of quiver gauge theories obtained as the
worldvolume theory on a stack of $N$ D3-branes placed at the conical
singularity of $\mathbb C^3/\Gamma$, paying particular attention to the case
where $\Gamma$ is non-Abelian.  This section also presents a
general construction of the set of permutations $\mathcal A$ and rephasing
symmetries $\mathcal B$ and $\mathcal C$. Section \ref{sec:string} describes
how the field theory results match those of the dual string theory states.
Section \ref{sec:examples} contains a number of explicit examples
motivating our claims.  Finally, in section \ref{conclusions}, we present
our conclusions and point out some open questions.
We relegate explicit details of the Abelianization of discrete subgroups
of $\mathrm{SU}(3)$ to Appendix~\ref{sec:abelian} and the representations
of $\Delta(27)$ to Appendix~\ref{sec:gtDelta27}.

\section{Discrete symmetries of orbifold quiver gauge
theories}\label{sec:nonabelian}

Before turning to the construction of the discrete symmetry operations
$\mathcal A$, $\mathcal B$ and $\mathcal C$, we first summarize the
basic features of $\mathcal N=1$ quiver gauge theories corresponding
to a stack of $N$ D3-branes at the singular point of the
$\mathbb C^3/\Gamma$ orbifold.  We are, of course, especially interested
in the case where $\Gamma$ is non-Abelian.  Pioneering work towards
constructing these orbifold theories began with
\cite{Douglas:1996sw,Douglas:1997de,Douglas:1997zj}, and subsequently the
quivers were more fully developed in \cite{Kachru:1998ys,Lawrence:1998ja}.

\subsection{$\mathcal N=1$ gauge theories from $\mathbb C^3/\Gamma$ orbifolds}

{}From our point of view, an
$\mathcal N=1$ quiver is essentially a set of nodes representing gauge groups and
links which describe chiral multiplets transforming in the bifundamental representation.
For a discrete orbifold group $\Gamma\subset\mathrm{SU}(3)$, the nodes
of the quiver may be placed in one-to-one correspondence with the
irreducible representations $\mathbf r_i$ of $\Gamma$, and the
corresponding gauge groups are taken to be $\mathrm{SU}(n_iN)$ where
$n_i=\dim(\mathbf r_i)$.  In order to obtain the bifundamental matter,
we must choose an appropriate embedding of the orbifold action in
$\mathbb C^3$, corresponding to a (possibly reducible) three-dimensional
representation $\mathbf3$ of $\Gamma$.  The number of bifundamentals
stretching from node $i$ to node $j$ is then given by $b_{ij}$ in
the decomposition
\begin{equation}
\mathbf 3\otimes\mathbf r_i = \oplus_j b_{ij}\mathbf r_j.
\label{eq:3tensori}
\end{equation}
{}From this definition, it is clear that $b_{ij}$ is related to the
familiar adjacency matrix via
\be
a_{ij}=b_{ij}-b_{ji}.
\ee
Although not directly encoded in the quiver diagram itself, the
(cubic) superpotential may be fully determined from the properties
of the orbifold group $\Gamma$.  In the notation of \cite{Lawrence:1998ja},
the superpotential is given by
\begin{equation}
W=\sum h_{ijk}^{f_{ij},f_{jk},f_{ki}}
\Tr\left(\Phi^{ij}_{f_{ij}}\Phi^{jk}_{f_{jk}}\Phi^{ki}_{f_{ki}}\right),
\label{eq:spot}
\end{equation}
where
\begin{equation}
h_{ijk}^{f_{ij},f_{jk},f_{ki}}=\epsilon_{\alpha\beta\gamma}
(Y_{f_{ij}})^\alpha_{v_i\overline v_j}
(Y_{f_{jk}})^\beta_{v_j\overline v_k}
(Y_{f_{ki}})^\gamma_{v_k\overline v_i},
\label{eq:hijk}
\end{equation}
and $(Y_{f_{ij}})^\alpha_{v_i\overline v_j}$ is the Clebsch-Gordan
coefficient corresponding to the decomposition of (\ref{eq:3tensori}).
Note that $f_{ij}=1,\ldots,b_{ij}$ labels the specific link showing up
in the above decomposition.

\subsection{Discrete symmetries from one dimensional representations of
$\Gamma$}

Given a quiver theory constructed as above, our goal is now to identify
potential discrete global symmetries of the quiver.  These symmetries
fall naturally into two categories:
\begin{enumerate}
\item Permutation symmetries which maps fields to fields and gauge groups
to gauge groups.  Motivated by the notation of \cite{Gukov:1998kn},
we label these permutations as $\mathcal A$ type symmetries.
\item Symmetries that rephase fields in such a way as to leave all trace
type gauge invariant operators inert.  These rephasing symmetries may be
thought of as anomaly free discrete subgroups of the global $\mathrm U(1)$
symmetries acting at each node of the quiver.  These symmetries will be
denoted as either $\mathcal B$ type if they do not commute with the
permutations or $\mathcal C$ type if they do.
\end{enumerate}

\subsubsection{Permutation symmetries}

For a given quiver, the $\mathcal A$ type permutation symmetries are
easy to visualize.  If we view the quiver as a directed graph, then
the $\mathcal A$ type symmetries correspond to the subgroup of the
automorphism group of the quiver that leaves the superpotential
invariant.  In order to ensure that this is a symmetry of the full
theory, it is important that the superpotential remains invariant.
However, since this superpotential data is not manifest from the
quiver diagram itself, additional input must be considered when
determining the group of permutations generated by $\mathcal A$.
As will be seen in the examples below, this group could be substantially
smaller than the full automorphism group of the quiver diagram.

In order to systematically construct the $\mathcal A$ type
symmetries, we first note that good permutations can only map
amongst gauge groups of the same rank.  Since each gauge group
(node in the quiver) is labeled by an irreducible representation
$\mathbf r_i$ of $\Gamma$, this indicates that good permutations
must only map irreducible representations of the same dimension
into each other.  Such transformations are in fact naturally
furnished by the one-dimensional representations of $\Gamma$,
which we denote by $\mathbf 1_\alpha$, where $\alpha$ labels the
particular representation.  The action of $\mathbf 1_\alpha$ on
the nodes of the quiver follows directly from the tensor
product map
\be
\mathbf1_\alpha\otimes{\bb{r}{i}}= \bbs{r}_{\alpha(i)}.
\label{groupMap}
\ee
Note that, so long as $\mathbf{r}_i$ is irreducible, then so is
$\mathbf{r}_{\alpha(i)}$.  This may easily be seen because multiplication
by the one-dimensional representation on the left does not affect how
one may or may not block diagonalize a given matrix representation.
As a result, the map (\ref{groupMap}) indeed gives rise to a good
permutation on the nodes of the quiver, taking node $i$ to node $\alpha(i)$.

What remains to be demonstrated is that the transformation induced by
multiplication with $\mathbf{1}_\alpha$ also yields a permutation of
the bifundamental links consistent with invariance of the superpotential.
To see that this is the case, we recall that all superpotential terms
in the orbifolded theory that are mapped to each other by the action
of $\mathbf{1}_\alpha$ descend from a single superpotential term
in the original un-orbifolded parent $\mathcal N=4$ theory.  This turns out to be
sufficient to guarantee that the above action of the one-dimensional
representations is a symmetry of the orbifold quiver.

To see this, we first examine the permutation of the
links induced by $\mathbf{1}_\alpha$.  Here we may simply follow the
decomposition rules for the tensor product
\begin{equation}
\mathbf3\otimes\mathbf r_{\alpha(i)}=\mathbf3\otimes(\mathbf1_\alpha
\otimes\mathbf r_i)=\mathbf1_\alpha\otimes(\mathbf3\otimes\mathbf r_i)
=\oplus_j b_{ij}\mathbf1_\alpha\otimes\mathbf r_j
=\oplus_j b_{ij}\mathbf r_{\alpha(j)},
\end{equation}
which demonstrates that the matrices $b_{ij}$ before and after the
transformation are related by
\begin{equation}
b_{\alpha(i)\alpha(j)}=b_{ij},
\end{equation}
and therefore so are the adjacency matrices.
Since $\alpha(i)$ is just the relabeling of node $i$, this adjacency
matrix relation is precisely what is desired for generating good
permutations of the links.

Of course, we must also ensure that the superpotential remains invariant
under the mapping induced by $\mathbf1_\alpha$.  By examining (\ref{eq:spot})
and (\ref{eq:hijk}), we see that this would be the case, so long as
\begin{equation}
h_{\alpha(i)\alpha(j)\alpha(k)}=h_{ijk}
\label{eq:hrel}
\end{equation}
(where we have suppressed the $f_{ij}$ labels for brevity).  To prove this,
we have to turn to the properties of the Clebsch-Gordan decomposition.
In a quantum mechanical notation, the Clebsch-Gordan coefficient
$Y^\alpha_{v_i\overline v_j}$ corresponding to the decomposition
$\mathbf3\otimes\mathbf r_i\to\mathbf r_j$
may be written as the matrix element
\begin{equation}
Y^\beta_{v_i\overline v_j}=\langle \mathbf3,\mathbf r_i;\beta,v_i|
\mathbf3,\mathbf r_i;\mathbf r_j,v_j\rangle.
\end{equation}
At the same time, the Clebsch-Gordan coefficients for the
(rather trivial) multiplication by the one-dimensional representation
$\mathbf1_\alpha$ given in (\ref{groupMap}) are given by a $3\times3$
unitary matrix
\begin{equation}
U_{v_i\overline v_{\alpha(i)}}
=\langle\mathbf1_\alpha,\mathbf r_i;1,v_i|\mathbf1_\alpha,\mathbf r_i;
\mathbf r_{\alpha(i)},v_{\alpha(i)}\rangle.
\end{equation}
(Unitarity can be seen because we take $\mathbf1_\alpha\otimes
\bar{\mathbf1}_\alpha=\mathbf1_0$ without any phases, where
$\bar{\mathbf1}_\alpha$ is the complex conjugate of $\mathbf1_\alpha$,
and $\mathbf1_0$ is the trivial representation.)  Inserting a
complete set of states, we then have
\begin{eqnarray}
Y^\beta_{v_{\alpha(i)}\overline v_{\alpha(j)}}
&=&\langle\mathbf3,\mathbf r_{\alpha(i)};\beta,v_{\alpha(i)}|
\mathbf3,\mathbf r_{\alpha(i)};\mathbf r_{\alpha(j)},v_{\alpha(j)}\rangle
\nonumber\\
&=&\langle\mathbf1_\alpha,\mathbf r_i;\mathbf r_{\alpha(i)},v_{\alpha(i)}|
\mathbf1_\alpha,\mathbf r_i;1,v_i\rangle
\langle\mathbf3,\mathbf r_i;\beta,v_i|\mathbf3,\mathbf r_i;\mathbf r_j,v_j
\rangle\nonumber\\
&&\qquad\times
\langle\mathbf1_\alpha,\mathbf r_j;1,v_j|\mathbf1_\alpha,\mathbf r_j;
\mathbf r_{\alpha(j)},v_{\alpha(j)}\rangle\nonumber\\
&=&U^*_{v_i\overline v_{\alpha(i)}}Y^\beta_{v_i\overline v_j}
U_{v_j\overline v_{\alpha(j)}}.
\end{eqnarray}
Since $U_{v_i\overline v_{\alpha(i)}}$ is unitary, the superpotential
relation (\ref{eq:hrel}) then follows directly from the definition
(\ref{eq:hijk}).  Thus we have now shown that permutations
generated by the map (\ref{groupMap}) are indeed good symmetries
of the quiver gauge theory.

Finally, before turning to the rephasing symmetries, we make the key
observation that the $\mathcal A$ type permutation symmetries do indeed
form a group.  The reason for this is simply that the one dimensional
representations of a finite group are all of the representations of the
Abelianization of that group, and the Kronecker product acts as group
multiplication for these representations.  We therefore conclude that
the group formed by $\mathcal A$ type permutations is given by
\begin{equation}
\{\mathcal A\}=\bar\Gamma\equiv\Gamma/[\Gamma,\Gamma].
\label{eq:asymm}
\end{equation}

\subsubsection{Rephasing symmetries}

The second class of symmetries that may arise in the quiver theory are
discrete (global) $\mathrm U(1)$ transformations acting on the nodes
of the quiver.  As discussed in \cite{Burrington:2006aw,Burrington:2006pu},
these rephasing symmetries may be constructed by assigning discrete
$\mathrm U(1)$ charges $q_i$ to node $i$.  The $\mathrm{SU}(n_iN)$
adjoint gauge multiplets are of course inert under this transformation.
However, the bifundamentals $\Phi^{ij}_{f_{ij}}$ stretching from the
$i$th node to the $j$th node pick up a phase
\begin{equation}
\Phi^{ij}_{f_{ij}}\to\omega^{q_i-q_j}\Phi^{ij}_{f_{ij}},
\end{equation}
where $\omega$ is a primitive $k$-th root of unity (to be determined
below from the anomaly cancellation conditions).

To ensure that the above transformation is a symmetry of the quiver
theory, we must demand that it both leaves the superpotential invariant
and that it is anomaly free.  Invariance of the superpotential is
of course automatically satisfied, so only the anomaly condition comes
in to restrict the discrete charges.  As demonstrated in
\cite{Burrington:2006pu}, vanishing of the chiral anomaly at the $i$-th
node demands that
\begin{equation}
\omega^{\sum_ja_{ij}\tilde q_j}=1,
\end{equation}
where we recall that $a_{ij}$ is the adjacency matrix, and
$\tilde q_i\equiv q_in_iN$ is the charge weighted by the rank of the
$i$-th gauge group.  Solutions to the above fall into two classes.
The first class consists of continuous global $\mathrm U(1)$ symmetries,
which happens whenever $\tilde q_i$ is a zero eigenvector of the adjacency
matrix.  The second class occurs whenever the components of the vector
$\sum_ja_{ij}\tilde q_j$ share a common divisor $k$ (which can be taken
to be integral by appropriate scaling of the charges).  In other words
\cite{Burrington:2006pu}
\begin{equation}
\label{EigenMod}
\sum_ja_{ij} \tilde q_j \equiv 0 \quad (\mbox{mod} \; k)\qquad
\hbox{for }k\in\mathbb Z.
\end{equation}
Given a set of charges $\tilde q_i$ solving the above equation, the
chiral multiplets are then rephased according to
\be
\Phi^{ij}_{f_{ij}}\rightarrow e^{\fft{2\pi i}{kN}\left(\frac{\tilde q_i}{n_i}
-\frac{\tilde q_j}{n_j}\right)}\Phi^{ij}_{f_{ij}}.
\ee
Note, however, that since a transformation in the center of
$\mathrm{SU}(n_iN)$ at node $i$ may be written as $\Phi_{f_{ij}}^{ij}
\to e^{2\pi i/n_iN}\Phi_{f_{ij}}^{ij}$, the weighted charges
$\tilde q_i$ are only well defined mod~$k$.  In addition, this demonstrates
that such a rephasing symmetry is an order $k$ element.

While (\ref{EigenMod}) provides the only condition on the rephasing
symmetries, it does not provide a constructive procedure for obtaining
the $\mathcal B$ type (non-central) and $\mathcal C$ type (central)
transformations.  Nevertheless,
in practice, it is not too difficult to search for and obtain a consistent
set of charge assignments yielding the appropriate discrete transformations.
This will be demonstrated below in the examples.  For now, we simply note
that, for an $\mathcal A$ type transformation generated by $\mathbf 1_\alpha$
and a $\mathcal B$ type transformation specified by the charges $\tilde q_i$,
we may evaluate their commutator expression
\begin{equation}
\mathcal A^{-1}\mathcal B^{-1}\mathcal A\mathcal B:\quad
\Phi^{ij}_{f_{ij}}\to
e^{\fft{2\pi i}{kN}\left(\fft{\tilde q_i-\tilde q_{\alpha(i)}}{n_i}
-\fft{\tilde q_j-\tilde q_{\alpha(j)}}{n_j}\right)}
\Phi^{ij}_{f_{ij}}.
\end{equation}
Identifying this with a central element $\mathcal C$ ({\it i.e.}~$\mathcal A
\mathcal B=\mathcal B\mathcal A\mathcal C$) shows that the $\mathrm U(1)$
charges $\tilde s_i$ corresponding to $\mathcal C$ are given by
\begin{equation}
\tilde s_i=\tilde q_i-\tilde q_{\alpha(i)}\quad(\mbox{mod}\;k).
\label{eq:ccharge}
\end{equation}

Evaluating the commutator of $\mathcal A$ and $\mathcal C$ then gives
rise to a potentially new symmetry element $\mathcal D$ with $\mathrm U(1)$
charges given by
\begin{eqnarray}
\tilde t_i&=&\tilde s_i-\tilde s_{\alpha(i)}\quad(\mbox{mod}\;k)\nonumber\\
&=&\tilde q_i-2\tilde q_{\alpha(i)}+\tilde q_{\alpha(\alpha(i))}.
\end{eqnarray}
In order for $\mathcal C$ to be central, we demand that $\mathcal D$
is gauge equivalent to the identity.  Perhaps the simplest way for
this to occur is for $\tilde t_i$ to vanish at each node.  This gives
rise to a sufficient condition for $\mathcal C$ to be central
\begin{equation}
2\tilde q_{\alpha(i)}=\tilde q_i+\tilde q_{\alpha(\alpha(i))}\quad
(\mbox{mod}\;k).
\label{eq:scond}
\end{equation}
This condition, along with the anomaly requirement (\ref{EigenMod}),
may be used as a guide for constructing the appropriate $\mathcal B$
type rephasing symmetries of the quiver.  We note, however, that
while (\ref{eq:scond}) is a sufficient condition, there are other
possibilities that make $\mathcal C$ central as well.  One case that
often shows up is when the nodes are all of the same rank.  In this
case, instead of demanding the vanishing of $\tilde t_i$, it is sufficient
to ensure that they have a common value
\begin{equation}
\tilde t_i=\tilde t_j\quad(\mbox{mod}\;k)\qquad\mbox{when all}\; n_i=n_j.
\label{eq:scond2}
\end{equation}
This case arises in particular for abelian orbifolds of the form
$\mathbb C^3/\mathbb Z_n$.

Note that the charge condition (\ref{eq:scond}) may be iterated to yield a
general solution for the charges
\begin{equation}
\tilde q_{\alpha^n(i)}=n\tilde q_{\alpha(i)}-(n-1)\tilde q_i\quad
(\mbox{mod}\;k),\qquad n=1,2,\ldots,
\label{eq:ssoln}
\end{equation}
in terms of only two charges $\tilde q_i$ and $\tilde q_{\alpha(i)}$.
Recalling that this rephasing is an order $k$ operation, we may set
$n=k$ in the above expression to obtain $k(\tilde q_{\alpha(i)}
-\tilde q_i)=0\ (\mbox{mod}\;k)$, which is clearly always satisfied
for integer charges $\tilde q_i$.  This ensures the consistency of
the charge requirement (\ref{eq:scond}).

\subsection{The Heisenberg group}

As indicated in (\ref{eq:asymm}), the $\mathcal A$ type permutation
symmetries close to form a group isomorphic to $\bar\Gamma$, the
Abelianization of the orbifold action $\Gamma$.  Ideally, we would
be able to demonstrate explicitly that the $\mathcal B$ type rephasing
symmetries would also form a group isomorphic to $\bar\Gamma$.
However, we have as yet been unable to show this in a general manner.
Nevertheless, in practice, and as indicated in the examples, given
the proper identification of the $\mathcal A$ symmetries, it is
straightforward to construct the appropriate $\mathcal B$ generators
consistent with (\ref{EigenMod}) and either (\ref{eq:scond}) or
(\ref{eq:scond2}).  (The $\mathcal C$ generators then follow as a
direct consequence of commuting $\mathcal A$ and $\mathcal B$.)

The construction of the $\mathcal B$ generators is guided by noting
that since $\{\mathcal A\}$ is Abelian, it necessarily decomposes into
a set of cyclic groups
\begin{equation}
\{\mathcal A\}=\bar\Gamma=\mathbb Z_{a^{\;}_1}\otimes Z_{a^{\;}_2}
\otimes\cdots.
\label{eq:adecomp}
\end{equation}
We may then focus on a single $\mathbb Z_{a^{\;}_i}$ group at a time.
Being Abelian, this group may be generated by a single element
({\it i.e.}~some particular $\mathbf1_\alpha$) which we denote by
$\mathcal A_i$, and which cycles the nodes of the quiver.  The set of
nodes of the quiver then fall into distinct orbits of $\mathcal A_i$.
In general, the rephasing generator $\mathcal B_i$ corresponding to
this $\mathcal A_i$ may be obtained by the linear charge assignment
given in (\ref{eq:ssoln}), while in some cases iteration of
(\ref{eq:scond2}) may be required.  In any case, the related central
element $\mathcal C_i$ is given by the charge assignment of
(\ref{eq:ccharge}).

In this way, the complete group of global symmetries of the quiver
is constructed as a direct product of individual Heisenberg groups
generated by the elements $\mathcal A_i\mathcal B_i=\mathcal B_i
\mathcal A_i\mathcal C_i$ (where $i$ labels the group, and is not
summed over).  In other words, given the decomposition (\ref{eq:adecomp}),
the discrete symmetry group takes the form
\begin{equation}
\mathrm{Heis}(\zet_{a^{\;}_1},\zet_{a^{\;}_1})\otimes
\mathrm{Heis}(\zet_{a^{\;}_2},\zet_{a^{\;}_2})\otimes\cdots,
\label{eq:mheis2}
\end{equation}
which is just the decomposition (\ref{eq:manyheis}) highlighted in the
introduction.

\section{String theory interpretation and torsion cycles}\label{sec:string}

The result (\ref{eq:mheis2}) takes on added physical significance when
the quiver gauge theory is related to the dual string picture.  The
general framework is of course clear: following the general ideas of
\cite{Gukov:1998kn} and \cite{Burrington:2006uu}, we wish to identify
the symmetry generators $\mathcal A$, $\mathcal B$ and $\mathcal C$
in the field theory with corresponding operators counting the number of
wrapped F-strings, D-strings and D3-branes, respectively, in the dual
string theory.  Some subtleties arise, however, in making precise the
field theory/string theory connection in cases were $\Gamma$ is
non-Abelian.  As a result, we find it worthwhile to make a distinction
between:
\begin{enumerate}
\item The orbifold quiver gauge theory.
\item The near horizon manifold $S^5/\Gamma$.
\item The string theory orbifold $\mathbb C^3/\Gamma$.
\end{enumerate}

In the first case, we have demonstrated that the orbifold quiver gauge
theory admits the finite Heisenberg group (\ref{eq:manyheis}) as a
group of global symmetries of the field theory.  Based on AdS/CFT,
this field theory ought to be dual to string theory on the horizon
manifold $S^5/\Gamma$.  To show that the symmetries match on both sides
of the duality, we need information on the homology classes of $S^5/\Gamma$.
After all, we expect the $\mathcal A$ transformations to be identified
with operators counting the number of F-strings in the dual string
theory.  The structure of eigenvalues of these operators is clearly
determined by the first homology class $H_1(S^5/\Gamma)$.  Furthermore,
based on S-duality, the $\mathcal B$ transformations may be associated
with operators counting the number of D-strings; these operators are
also valued in $H_1(S^5/\Gamma)$.  Finally, the $\mathcal C$ transformations
match the operators counting wrapped D3-branes and is determined by
the third homology class, $H_3(S^5/\Gamma)$.  While $S^5$ itself has no
non-trivial cycles, orbifolds of $S^5$ may admit torsion one and three-cycles,
which are exactly what is required to allow this duality to work.

There is a slight subtlety in the identification of the operators of
$\mathcal C$. Namely, it may seem curious that the D3-brane number operator
is somehow related to the F-string and D-string number operators.  However,
we note that Poincar\'e duality relates the torsion free part of homology
groups by $(H_p)_{TF}=(H_{(d-p)})_{TF}$, while the (cyclic) torsion parts
of the homology are related by $\mathrm{Tors}(H_p)= \mathrm{Tors}(H_{(d-p-1)})$.
We are in particular interested in the torsion parts; on the five sphere,
$\mathrm{Tors}(H_1)= \mathrm{Tors}(H_{(5-1-1)})= \mathrm{Tors}(H_{(3)})$
(see for example \cite{Hatcher}).  Therefore, the torsion cycles that
F-strings or D-strings may wrap are isomorphic to torsion cycles that
D3-branes may wrap. As was shown explicitly in \cite{Morrison:1998cs}, in
the more general case of branes placed in generic toric singularities,
this is no longer the case, and one can have that $H_1(H)$ is not isomorphic
to $H_3(H)$ where $H$ is the near horizon space.

For $S^5/\Gamma$, on the other hand, the non-trivial homology is given by
(\ref{eq:homo}).  As a result, the dual picture of F-strings and D-strings
wrapping torsion cycles gives rise to the identical Heisenberg group
(\ref{eq:manyheis}) that was obtained in the field theory analysis.
This statement is the extension of \cite{Gukov:1998kn,Burrington:2006uu}
to the non-Abelian case.

Although the duality between the symmetries of the quiver gauge theory
and the near horizon manifold $S^5/\Gamma$ is clear, the connection
with string theory on $\mathbb C^3/\Gamma$ is less so (at least in
the non-Abelian case).  This is because a rigorous understanding of
D3-branes near a non-Abelian orbifold singularity involves the
generalization of \cite{Douglas:1996sw,Douglas:1997de,Douglas:1997zj}
to the non-Abelian case, and this is as yet incomplete.  Here we
simply make an observation on the structure of twisted sectors in
string theory on orbifolds. In string theory we are mainly familiar
with global orbifolds by Abelian groups such as $\mathbb{Z}_N$ or
$\mathbb{Z}_N \times \mathbb{Z}_N$.  In this case, the space of twisted
sectors is in correspondence with elements of the group excluding the
identity element.  In general, however, twisted sectors are in correspondence
with conjugacy classes \cite{Dixon:1986jc}.  Consider a string field $X$
whose boundary conditions are twisted by an element $g\in \Gamma$, namely
$X(\sigma + 2\pi)=gX(\sigma)$. For any element $h\in \Gamma$ we can
act on the previous relation on the left: $h X(\sigma + 2\pi)=hgX(\sigma)
=(hgh^{-1}) \, hX(\sigma)$. This means that strings twisted by
$hgh^{-1}$ are all in the same sector. Thus twisted sectors are defined
only up to conjugacy classes.  That is, there is one sector per conjugacy
class in $\Gamma$.

Since the field theory states are classified by the
Abelianization of $\Gamma$, to make a connection between the quiver
and orbifold pictures, we must appropriately relate the conjugacy classes
of $\Gamma$ with the Abelianization of $\Gamma$.  In the non-Abelian
case, however, this relation is not so clear.  Moreover, unlike the
Abelianization of a group, which itself is a group, there is no
natural group structure on conjugacy classes.  Nevertheless, we fully
expect that (\ref{eq:manyheis}) properly describes the global symmetries
of the theory on both the gauge theory and the string theory sides of
the duality.

\section{Examples}\label{sec:examples}

As indicated above, we are interested in $\mathcal N=1$ quiver theories
that may be obtained by orbifolding $\mathcal N=4$ super-Yang Mills by
a group $\Gamma$.  To ensure $\mathcal N=1$ supersymmetry, $\Gamma$
must be restricted to be a discrete subgroup of $\mathrm{SU}(3)$.  In
fact, all such discrete subgroups have been classified
\cite{Blichfeldt}, and many of the resulting quiver gauge theories
have been described in \cite{Hanany:1998sd,Feng:2000af,Feng:2000mw}.

We follow the discussion of \cite{Hanany:1998sd} concerning
the non-Abelian discrete subgroups of $\mathrm{SU}(3)$.  The
relevant subgroups which are not
contained in $\mathrm{SU}(2)$ fall into two infinite series,
$\Delta(3n^2)$ and $\Delta(6n^2)$, as well as the exceptional subgroups
$\Sigma(36)$, $\Sigma(60)$, $\Sigma(72)$, $\Sigma(168)$, $\Sigma(216)$,
$\Sigma(360)$ and $\Sigma(36\times3)$, $\Sigma(60\times3)$,
$\Sigma(168\times3)$, $\Sigma(216\times3)$, $\Sigma(360\times3)$.  We
note that the subgroups in the infinite series are
subgroups of $\mathrm{SU}(3)$ for $n=0\mod3$, and $\mathrm{SU}(3)/\mathbb Z_3$
for $n\ne0\mod3$.  Similarly, the latter set of exceptional subgroups are
subgroups of $\mathrm{SU}(3)$, while the former set are subgroups of
$\mathrm{SU}(3)/\mathbb Z_3$ \cite{Hanany:1998sd}.

\begin{table}[t]
\begin{center}
\begin{tabular}{l|l}
$\Gamma$&$\bar\Gamma$\\
\hline
$\Delta(3n^2)$&$\mathbb Z_3\times\mathbb Z_3\hphantom{\mathbb Z_3}$
for $n=0\mod3$\\
&$\mathbb Z_3\hphantom{Z_3\times\mathbb Z_3}$ for $n\ne0\mod3$\\
$\Delta(6n^2)$&$\mathbb Z_2$\\
$\Sigma(36)$&$\mathbb Z_4$\\
$\Sigma(36\times3)$&$\mathbb Z_4$
\end{tabular}
\end{center}
\caption{Some subgroups $\Gamma$ of $\mathrm{SU}(3)$ and their
Abelianization $\bar\Gamma$.}
\label{tbl:abel}
\end{table}

In Appendix~\ref{sec:abelian}, we compute the Abelianization of
several of these groups.  The results are given in Table~\ref{tbl:abel}.
Note that the Abelianization of $\Gamma$ is not necessarily related
to the center of $\mathrm{SU}(3)$, as groups such as $\mathbb Z_2$
and $\mathbb Z_4$ may arise for $\bar\Gamma$.  In most cases,
$\bar\Gamma$ is given by a single cyclic group $\mathbb Z_k$, in which
case the discrete symmetry group of the quiver is a single copy of
$\mathrm{Heis}(\mathbb Z_k\times\mathbb Z_k)$.  The case of
$\Delta(3n^2)$ for $n=0\mod3$ is somewhat interesting, though, as its
Abelianization contains two cyclic factors.  We will highlight this
case below by considering $\Delta(27)$.  However, we start with the
familiar example of the $\mathbb Z_3$ orbifold in order to introduce the
language of discrete symmetries constructed from one-dimensional
representations of $\Gamma$.

\subsection{The $\zet_3$ quiver}

The discrete symmetries of the $\mathbb Z_3$ quiver formed the
basis of the analysis of Gukov, Rangamani and Witten in \cite{Gukov:1998kn}.
Although this group is Abelian, we find this example instructive as
it enables us to emphasize the r\^ole of one-dimensional representations
as generators of the $\mathcal A$ type permutation symmetries.

The cyclic group $\zet_3$ has only one generator, which we call $A$,
and which satisfies
\be
A^3=1.
\ee
Since $\zet_3$ is Abelian, it only has one-dimensional representations.
For the same reason, the conjugacy classes are given by just the individual
group elements.  The character table is then
\begin{equation}
\begin{tabular}{c||c|c|c}
& 1 & $A$ & $A^2$ \\
\hline \hline
$\bbs{1}_0$ & $1$ & $1$ & $1$ \\
\hline
$\bbs{1}_1$ & 1 & $\gamma$ & $\gamma^2$ \\
\hline
$\bbs{1}_2$ & 1 & $\gamma^2$ & $\gamma$ \\
\end{tabular}
\end{equation}
where $\gamma$ is a cube root of unity.

\begin{figure}[t]
\begin{center}
\input{z3.pstex_t}
\end{center}
\caption{The $\mathbb Z_3$ quiver.}
\label{fig:z3}
\end{figure}
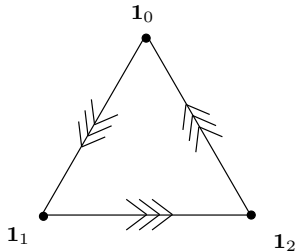

To construct the $\mathbb Z_3$ quiver, we must specify how it acts on
the space $\mathbb C^3$.  This corresponds to specifying an appropriate
faithful three dimensional representation to act on the global index,
which is also required to be a subgroup of $\mathrm{SU}(3)$.
We take this to be
\be
A=\begin{pmatrix}
\gamma & 0 & 0 \\
0 & \gamma & 0 \\
0 & 0 & \gamma
\end{pmatrix}.
\ee
Now it is easy to see what happens.  The three dimensional representation
is a reducible representation:
$\bbs{3}=\bbs{1}_1 \oplus \bbs{1}_1 \oplus \bbs{1}_1$.
It is three copies of the same representation, and so in the quiver
we expect three arrows all pointing in the same direction from a given node.
Also, we expect a global $\mathrm{SU}(3)$ symmetry because one may recombine
like representations into each other.  We will check this at the end of the
calculation%
\footnote{The global $\mathrm{SU}(3)$ symmetry is a statement that there are
three identical representations acting on the global index, and this remains as
a global symmetry.  This is analogous to the unbroken gauge factors.
Recall that in the regular representation, an $n$-dimensional representation
appears $n$ times, and it is this fact that leads to the unbroken
$\mathrm U(n)$ gauge group.  In the holographic limit, such symmetries become
$\mathrm{SU}(n N)$.}.
The structure of the quiver is given by
\be
\bbs{3}\otimes \bbs{1}_i=\bbs{1}_{i+1}+\bbs{1}_{i+1}+\bbs{1}_{i+1}.
\ee
Thus the quiver, shown in Fig.~\ref{fig:z3}, has three arrows pointing
from one node to the next in the cyclic order
$\bbs{1}_0\rightarrow\bbs{1}_1\rightarrow\bbs{1}_2$.
We label the bifundamental fields by $C_{a,b}^i$ where $a$ ($b$) labels the
node that the arrow points from (to), and $i$ is a global $\mathrm{SU}(3)$
index labeling which arrow of the three that is being referred to, and
which directly corresponds to the global index in the
$\mathcal{N}=4$ theory from which the above orbifold theory descends.

To compute the superpotential, we note that the Clebsch-Gordan coefficients
are all trivial (taken to be unity in our conventions), as there is only
one way of combining one-dimensional representations.  In this case, the
superpotential is easy to deduce, and is simply
$W=\epsilon_{ijk} C_{01}^i C_{12}^j C_{20}^k$.  As claimed, there is a
global $\mathrm{SU}(3)$ symmetry which directly corresponds to mixing the
three irreducible representations in the $\bbs{3}$ (reducible) representation.

Given the quiver and the superpotential, we are now in a position to
highlight the global symmetries $\mathcal A$, $\mathcal B$ and $\mathcal C$.
Starting with the $\mathcal A$ type shift symmetry, we recall that it
is generated by the action of the one-dimensional representations on
the nodes and links of the quiver according to (\ref{groupMap}).
Taking $\bbs{1}_{1}$ to generate the $\mathcal A$ symmetry, we see that
it simply maps the nodes cyclicly
$\bbs{1}_0\rightarrow\bbs{1}_1\rightarrow\bbs{1}_2$.  The mapping of the
fields is also easy, as all Clebsch-Gordan coefficients are trivial,
and is given simply by
\begin{eqnarray}
\mathcal A:&&C_{a,b}^i\rightarrow C_{a+1, b+1}^i
\label{eq:z3a}
\end{eqnarray}
(where the node labels are taken mod~3).  This is clearly a symmetry of the
superpotential.

To obtain the $\mathcal B$ and $\mathcal C$ rephasing symmetries, we start
with the adjacency matrix
\be
\bbs{a}=
\begin{pmatrix}
0 & 3 & -3 \\
-3 & 0 & 3 \\
3 & -3 & 0
\end{pmatrix}.
\ee
The anomaly equation (\ref{EigenMod}) takes the form
\be
\bbs{a}\cdot\bbs{v}\equiv 0 \quad (\mbox{mod 3}),
\ee
and can be satisfied by any vector $\bbs{v}$ of integers.  Since all
gauge groups have the same rank, an appropriate choice of charge
vector satisfying (\ref{eq:scond2}) with $\tilde t_i=1\mod3$ is given by
$\bbs{v}=(0,0,1)$.  On the fields, this corresponds to
a rephasing symmetry
\bea
\label{eq:z3b}
\mathcal{B}:
&&C^i_{0,1}\rightarrow C^i_{0,1}, \nn \\
&&C^i_{1,2}\rightarrow \omega^{-1} C^i_{1,2}, \\
&&C^i_{2,0}\rightarrow \omega^{1} C^i_{1,2}, \nn
\eea
where $\omega=\exp(2\pi i/(3N))$.  Note that
this symmetry applied three times gives a member of the
center of the gauge group, and so is gauge equivalent to
the identity.

Clearly $\mathcal{A}$ and $\mathcal{B}$ do not commute.
They in fact close on the rephasing symmetry given by the
vector $\bbs{v}'=(1,0,-1)$, acting on the fields as
\bea
\label{eq:z3c}
\mathcal{C}:&&C^i_{0,1}\rightarrow \omega C^i_{0,1}, \nn \\
&&C^i_{1,2}\rightarrow \omega C^i_{1,2}, \\
&&C^i_{2,0}\rightarrow \omega^{-2} C^i_{1,2}. \nn
\eea
Although it is not obvious here, $\mathcal{A}$ and $\mathcal{C}$
commute (up to a gauge transformation).  This is because the charges needed
to close the $\mathcal{AC}$ commutator are $\bbs{v}''=(-2,1,1)$, which
rephases in the exact same way as $\bbs{v}''=(-3,0,0)$.  The latter is in the
center of the gauge group because all charges are divisible by three.
Examination of (\ref{eq:z3a}), (\ref{eq:z3b}) and (\ref{eq:z3c}) indicates
that the symmetry generators obey the relations
\bea
&&\mathcal{A}^3=\mathcal{B}^3=\mathcal{C}^3=1, \quad \mathcal{AB}=\mathcal{BAC}, \nn\\
&& \mathcal{AC}=\mathcal{CA},\quad \mathcal{BC}=\mathcal{CB},
\eea
up to the center of the gauge group \cite{Gukov:1998kn}.  As expected,
this is just the finite Heisenberg group
$\mathrm{Heis}(\mathbb Z_3\times\mathbb Z_3)$.

\subsection{The $\Delta(27)$ quiver}

This particular orbifold has been much studied in the literature, and
is one of the simplest examples of a non-Abelian orbifold
\cite{Hanany:1998sd,Muto:1998na,Muto:1999fv,Muto:1999zw,Feng:2000af,Feng:2000mw,Verlinde:2005jr}.
Here we will try to be as detailed as possible, so that the general
structure of the discrete symmetries becomes apparent in this example.

The group theory details for $\Delta(27)$ are given in
Appendix~\ref{sec:gtDelta27}.  Here we note that, for any $n=0 \mod 3$,
the group $\Delta(3n^2)$ is a subgroup of the full $\mathrm{SU}(3)$,
and has $n^2/3-1$ three-dimensional representations and 9 one-dimensional
representations. The three-dimensional representations may be labeled by
integer pairs $i,j$ with $0\le i,j\le n$, along with the equivalence
$\mathbf3_{i,j}=\mathbf3_{j,-i-j}=\mathbf3_{-i-j,i}$ (where labels are
taken mod~$n$).  The $\mathbf3_{0,0}$, $\mathbf3_{n/3,n/3}$ and
$\mathbf3_{2n/3,2n/3}$ representations are reducible, and fall apart into
the nine one-dimensional representations, which we may label as
$\mathbf1_{i,j}$ as follows:
\begin{equation}
\mathbf3_{i\times n/3,i\times n/3}\quad\to\quad \mathbf1_{i,0},
\mathbf1_{i,1},\mathbf1_{i,2}.
\end{equation}
In the present case of $n=3$, we only have two three-dimensional
representations (which we therefore label $\bbs{3}$ and $\bbbs{3}$).  Using
the $\bbs{3}$ representation as the faithful representation that acts on the
global symmetry index, one may easily deduce the structure of the quiver
(see the table of tensor products in Appendix~\ref{sec:gtDelta27}).
This is given in Fig.~\ref{fig:d27}.
We label the fields pointing from the $\bb1{i,j}$ node as $A_{i,j}$ and
those pointing to the $\bb1{i,j}$ as $B_{i,j}$.  In addition, we label
the three other fields (pointing from the $\mathbf3$ to the $\bar{\mathbf3}$)
as $C_{i}$, $i=0,1,2$.

\begin{figure}[t]
\begin{center}
\input{delta27.pstex_t}
\end{center}
\caption{The $\Delta(27)$ quiver.}
\label{fig:d27}
\end{figure}
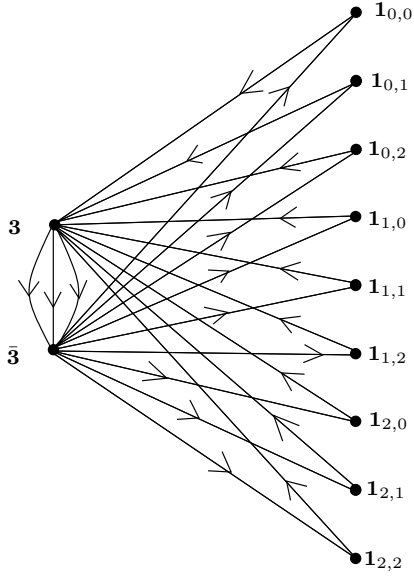
\begin{table}[t]
\begin{center}
\begin{tabular}{l|lll}
$\alpha_{i,j,k}$&$C_0$&$C_1$&$C_2$\\
\hline
$B_{0,j}A_{0,j}$&$0$&$-1$&$1$\\
&$0$&$-1$&$\gamma^2$\\
&$0$&$-1$&$\gamma$\\
\hline
$B_{1,j}A_{1,j}$&$1$&$0$&$-1$\\
&$\gamma^2 $&$0$&$-1$\\
&$\gamma $&$0$&$-1$\\
\hline
$B_{2,j}A_{2,j}$&$-1$&$1$&$0$\\
&$-1$&$\gamma^2$&$0$\\
&$-1$&$\gamma$&$0$
\end{tabular}
\end{center}
\caption{The superpotential coefficients $\alpha_{i,j,k}$ multiplying
$B_{i,j}A_{i,j}C_k$.  Here $\gamma=e^{2\pi i/3}$ is a cube root of unity.}
\label{tbl:abc}
\end{table}

The superpotential may be obtained using the Clebsch-Gordan coefficients
given in Appendix~\ref{sec:gtDelta27}.  The result is
\begin{equation}
W=\alpha_{i,j,k}B_{i,j}A_{i,j}C_k,
\end{equation}
with the coefficients $\alpha_{i,j,k}$ given in Table~\ref{tbl:abc}.
More succinctly, we have
\begin{equation}
\alpha_{i,j,k}=\delta_{i-1,k}\gamma^{-j}-\delta_{i+1,k},
\end{equation}
where all indices are to be taken modulo 3, and where $\gamma^3=1$.

To construct the shift
symmetries, we may begin with the $\mathbb Z_3$ symmetry generated by
$\mathbf1_{1,0}$ which takes $\bb1{i,j} \to \bb1{i+1,j}$ as well as
$\mathbf3\to\mathbf3$ and $\bar{\mathbf3}\to\bar{\mathbf3}$.
Although the $\mathbf3$ and $\bar{\mathbf3}$ nodes are inert under
this transformation, the $C_i$ links are permuted.
The action of this symmetry , which we denote $\mathcal A$, on the
bifundamentals is thus
\begin{equation}
{\mathcal A} :\quad A_{i,j}\to A_{i+1,j},\qquad
B_{i,j}\to B_{i+1,j},\qquad C_k\to C_{k+1}.
\label{eq:z31}
\end{equation}
The Abelianization of $\Delta(27)$ contains a second $\mathbb Z_3$ factor, and
so we expect a second $\mathbb Z_3$ symmetry of the quiver.  Noting
that the nine one-dimensional representations form the group
$\mathbb Z_3\times\mathbb Z_3$ under ordinary multiplication
($\mathbf1_{i,j}\times\mathbf1_{k,l}=\mathbf1_{i+k,j+l}$), we may take
this second $\mathbb Z_3$ to be generated by the action of $\mathbf1_{0,1}$.
This acts on the second index of the $B_{i,j}$ and $A_{i,j}$ fields.
However, this is accompanied by additional rephasings as well.  The action
of this $\mathbb Z_3'$, which we call $\mathcal A'$, is%
\footnote{Equivalently, one may choose to rephase $A_{i,j}$ and leave
$B_{i,j}$ alone. This symmetry is pure gauge, given by simultaneously using
the center of the gauge groups $\bbs{3}$ and ${\bbbs{3}}$, hence leaving
the $C$ fields unchanged.}
\begin{equation}
{\mathcal A}':\quad A_{i,j}\to A_{i,j+1},\qquad
B_{i,j}\to \gamma^{i}B_{i,j+1},\qquad C_k\to\gamma^{1-k} C_k.
\label{eq:z32}
\end{equation}
We note that $\mathbb Z_3$ and $\mathbb Z_3'$ in (\ref{eq:z31})
and (\ref{eq:z32}) do not strictly commute.  However, they commute up to
a rephasing of the fields
\begin{equation}
\quad A_{i,j}\to A_{i,j},\qquad
B_{i,j}\to \gamma B_{i,j},\qquad C_k\to\gamma^2 C_k,
\end{equation}
which is in the center of the gauge group associated
with the $\bbbs{3}$ representation.  This is because $\gamma$ is a third root
of unity, and so is also a $3N$-th root of unity, thus ensuring that the
above rephasing is in the center of the $\mathrm{SU}(3N)$ gauge group
associated with the $\bbbs{3}$ node.  Thus the $\Delta(27)$ quiver does
in fact admit
a $\mathbb Z_3\times\mathbb Z_3$ symmetry.  Incidentally, we note that the
quiver diagram shown in Fig.~\ref{fig:d27} actually admits an $S_9$ symmetry
permuting the nine `singlet' nodes.  The superpotential, however, is only
invariant under the $\mathbb Z_3\times\mathbb Z_3$ subgroup of the full
$S_9$ permutation group.

We now turn to the rephasing symmetries.  Since the $\mathcal A$ type
symmetries generate $\mathbb Z_3\times\mathbb Z_3$, we expect the
$\mathcal B$ type rephasings to generate this identical group.
These rephasings are relatively easy to deduce from the
results for the $\mathbb Z_3$ orbifold theory \cite{Gukov:1998kn} given
above.  There, the appropriate rephasings were given by the charges
$(0,0,1)$ on the three nodes of the quiver.  Therefore, by extension,
we find the charge assignments for the nine $\mathrm{SU}(N)$ nodes to
be correctly given by $(0,0,0,0,0,0,1,1,1)$ for what we call $\mathcal B$,
and $(0,0,1,0,0,1,0,0,1)$ for what we call $\mathcal B'$ (the $\mathrm{SU}(3N)$
nodes are assigned zero charge).  This generates the rephasings
\begin{eqnarray}
{\mathcal B}:&&\quad A_{i,j}\to\omega A_{i,j},\qquad
B_{i,j}\to\omega^{-1} B_{i,j},\quad i=2\hbox{ only},\\
{\mathcal B'}:&&\quad A_{i,j}\to\omega A_{i,j},\qquad
B_{i,j}\to\omega^{-1} B_{i,j},\quad j=2\hbox{ only},
\end{eqnarray}
where the other fields are not rephased.  Here $\omega=\exp(2\pi i/(3N))$
is a primitive $3N$-th root of unity.

It is clear that the primed symmetries commute with unprimed symmetries.
However, $\mathcal A$ and $\mathcal B$ do not commute, and $\mathcal A'$
and $\mathcal B'$ do not commute.  Instead, they commute up to the respective
symmetries
\begin{eqnarray}
{\mathcal C}:&&\quad A_{i,j}\to\omega^{i-1} A_{i,j},
\quad B_{i,j}\to\omega^{-(i-1)} B_{i,j}, \\
{\mathcal C'}:&&\quad A_{i,j}\to\omega^{j-1} A_{i,j},
\quad B_{i,j}\to\omega^{-(j-1)} B_{i,j}.
\end{eqnarray}
Again, it is clear that primed and unprimed symmetries commute.  It is also
clear that ${\mathcal B}$ and $\mathcal C$ (primed or not) commute.  Finally,
$\mathcal A$ and $\mathcal C$ also commute, but only up to the center of the
gauge group.  In particular, the member of the center of the gauge group that
closes the $\mathcal{AC}$ commutator is
\begin{eqnarray}
&& A_{i,j}\to\omega^{-2} A_{i,j},\quad B_{i,j}\to\omega^{2} B_{i,j},
\quad i=0,\nonumber \\
&& A_{i,j}\to\omega A_{i,j},\quad B_{i,j}\to\omega^{-1} B_{i,j},\quad i\neq0,
\end{eqnarray}
and the member of the center of the gauge group that closes the
$\mathcal{A'C'}$ commutator is
\begin{eqnarray}
&&A_{i,j}\to\omega^{-2} A_{i,j},\quad B_{i,j}\to\omega^{2} B_{i,j},
\quad j=0,\nonumber \\
&&A_{i,j}\to\omega A_{i,j},\quad B_{i,j}\to\omega^{-1} B_{i,j},\quad j\neq0,
\end{eqnarray}
where in both cases the $C_k$ are unchanged.  These two rephasings can be
seen to correspond to the center of the gauge groups by assigning charge
$-3$ to each of the $\mathrm{SU}(3N)$ nodes and then assigning
$(3,3,3,0,0,0,0,0,0)$ and $(3,0,0,3,0,0,3,0,0)$ charge vectors to the
$\mathrm{SU}(N)$ nodes, respectively.

Thus the structure of the global symmetries is that each $\mathbb Z_3$ shift
symmetry ($\mathcal A$ or $\mathcal A'$) associated with a one-dimensional
representation has a corresponding rephasing $\mathbb Z_3$ symmetry
($\mathcal B$ or $\mathcal B'$).  These two symmetries close on a final
$\mathbb Z_3$ symmetry ($\mathcal C$ or $\mathcal C'$).  All generators
are of order three (up to the center of the gauge group), and the primed
and unprimed ones commute.  As a result, taken together, they form a
direct product of two Heisenberg groups
\begin{equation}
\mathrm{Heis}(\mathbb Z_3 \times \mathbb Z_3) \times
\mathrm{Heis}(\mathbb Z_3 \times \mathbb Z_3),
\end{equation}
in agreement with the expectation from (\ref{eq:manyheis}), where we
note that the Abelianization of $\Delta(27)$ is just $\mathbb Z_3\times
\mathbb Z_3$.

\section{Conclusions}\label{conclusions}

Our work demonstrates that quiver gauge theories obtained as worldvolume theories
on a stack of  $N$ D3-branes on the singular point of
$\mathbb{C}^3/\Gamma$ where $\Gamma$ is non-Abelian admit a discrete group
of global symmetries which may be expressed as a product of Heisenberg groups.
It is worth mentioning, however, that the actual matching which we
perform is with string theory on the near horizon manifold
AdS$_5\times S^5/\Gamma$.  This highlights the importance of the decoupling
limit which is accompanied with the decoupling of various $\mathrm U(1)$'s,
some of them anomalous. We believe a similar structure should exist in the
general case of quiver gauge theories obtained as worldvolume theories on a
stack of $N$ D3-branes placed at the singular point of a toric variety which
can be obtained as a non-Abelian orbifold of some other toric variety.

More generally, our series of papers
\cite{Burrington:2006uu,Burrington:2006aw,Burrington:2006pu}
suggests that field theoretical methods might help in answering in full
generality the question of the spectrum of D-brane charges in string
theories on curved backgrounds and with fluxes whenever they admit field
theory duals. It has clearly been established that the Heisenberg group
structure is present in a variety of situations, including cascading
theories and Seiberg dual phases.  The larger question towards which our
work points is the computation of the spectrum on branes using AdS/CFT,
therefore predicting the outcome of the corresponding generalized cohomology
theory classifying the D-brane charges in string theory. For example, a
question arises motivated by a recent comment made in \cite{Evslin:2006cj}
about twisted K-theory being able to classify universality classes of
baryonic vacua in the Klebanov-Strassler background.  It is worth point
out that the argument of \cite{Evslin:2006cj} is entirely based on the
geometry of fluxes in the supergravity background.

Given our previous
work with cascading quiver gauge theories \cite{Burrington:2006aw}, it is
natural to suspect that field theoretic methods along the lines described
here should be relevant to understanding the spectrum of D-brane charges in
those backgrounds. Note that, to a large extent, the states charged under
the symmetries $\mathcal A$, $\mathcal B$ and $\mathcal C$ are determinant
operators in the field theory, also referred to as baryonic.  Here we would
also like to stress that while we have considered certain global symmetries,
we have not considered all possible symmetries, and a full classification and
their dual interpretation would be enlightening.  For example, the continuous
$U(1)$ baryon number symmetries (associated with various fractional branes)
do not commute with the shift operators, and close on other baryon $U(1)$
symmetries.  We hope to return to some of these issues in the future.

\section*{Acknowledgments}

We are thankful to P. de Medeiros, Y.-H. He, C. Herzog, I. Kriz,
D. Morrison and B. Uribe for various comments about mathematical aspects
of our work.  JTL is especially grateful to S. Teraguchi for raising
the issue of non-Abelian orbifolds and for suggesting that the previous
work on discrete symmetries may be extended to cover more general orbifold
constructions.  This work is supported in part by the US Department of
Energy under grant DE-FG02-95ER40899, the National Science Foundation
under grant PHY99-07949, the Israel Science Foundation, and the German-Israeli
Project (DIP) program (H.52).  JTL wishes to acknowledge the hospitality of
the Department of Physics at National Taiwan University and the Taiwan
National Center for Theoretical Sciences where this project was initiated,
and the Kavli Institute for Theoretical Physics where portions of this work
were completed.

\appendix

\section{The Abelianization of $G$}
\label{sec:abelian}

We have seen that the Abelianization of the orbifold group $\Gamma$
(which we denote $\bar\Gamma\equiv\Gamma/[\Gamma,\Gamma]$) plays a
key r\^ole in understanding the discrete global symmetries of the
orbifold quiver gauge theory.  In particular, for a quiver constructed
as an orbifold of $\mathcal N=4$ super-Yang Mills by $\Gamma$ via the
process in \cite{Lawrence:1998ja}, $\bar\Gamma$ (which is isomorphic
to the group formed by the one-dimensional representations of $\Gamma$
under ordinary multiplication) generates a set of permutation symmetries
mapping fields to fields and gauge groups to gauge groups.  We
expect this permutation mapping to work for non-supersymmetric orbifolds
as well as supersymmetric ones%
\footnote{In the non-supersymmetric case, one simply writes two different
kinds of arrows in the quiver: one type for fermions, and one type for bosons.
Since the gauge factors for both the fermions and bosons are mapped in the
same way, one still expects the one-dimensional representations to generate
symmetries of the resulting theory, again because the regular representation
is used on the gauge indices.  One can similarly argue that the (non-super)
potential that arises in these cases is still preserved by these symmetries.
It is interesting to see that this structure holds even for non-supersymmetric
orbifolds.}.
However here we restrict ourselves to the supersymmetric case.  In this
case, $\Gamma$ is a discrete subgroup of $\mathrm{SU}(3)$, and the
categorization of these subgroups is known \cite{Blichfeldt}.  In
this Appendix, we compute the Abelianization of $\Gamma$ for
the following examples:
\begin{equation}
\Delta(3n^2),\qquad\Delta(6n^2),\qquad\Sigma(36),\qquad\Sigma(36\times3).
\end{equation}

\subsection{The case $\Gamma=\Delta(3n^2)$}

The group $\Delta(3n^2)$ is generated by
\begin{equation}
A^3=B^n=C^n=1,
\end{equation}
and
\begin{equation}
BA=AC^{-1},\qquad CA=ABC^{-1}
\end{equation}
(with all other generators commuting).  As a result, the commutator subgroup
is generated by the elements $\{B^{-1}C^{-1},BC^{-2}\}$.  This is equivalent
to taking the generators
\begin{equation}
\{BC,C^3\}\quad\hbox{for}\quad n=0\mod3;\qquad
\{BC,C\}\quad\hbox{for}\quad n\ne0\mod3.
\end{equation}
Since a generic $\Delta(3n^2)$ element can be written as
\begin{equation}
g=A^aB^bC^c=A^a(BC)^bC^{c-b},
\end{equation}
it is easy to see that the Abelianization of $\Gamma=\Delta(3n^2)$ is
\begin{equation}
\bar\Gamma=\begin{cases}
\mathbb Z_3\times\mathbb Z_3&\hbox{if }n=0\mod3,\cr
\mathbb Z_3&\hbox{if }n\ne0\mod3.\end{cases}
\label{eq:z3z3}
\end{equation}

\subsection{The case $\Gamma=\Delta(6n^2)$}

The group $\Delta(6n^2)$ is generated by
\begin{equation}
A_1^2=A_2^3=B^n=C^n,
\end{equation}
with
\begin{eqnarray}
&&A_2A_1=A_1A_2^{-1},\qquad BA_1=A_1C,\qquad CA_1=A_1B,\nonumber\\
&&BA_2=A_2C^{-1},\qquad CA_2=A_2BC^{-1}.
\end{eqnarray}
Note that $\{A_1,A_2\}$ generate the permutation group $S_3$.  Thus
$\Delta(6n^2)$ is isomorphic to $(\mathbb Z_n\times\mathbb Z_n)\ltimes
S_3$.  A generic element of $\Delta(6n^2)$ may be written as
\begin{equation}
g=A_1^{a_1}A_2^{a_2}B^bC^c,
\end{equation}
and the commutator subgroup is generated by
\begin{equation}
\{A_2,B,C\}.
\end{equation}
As a result, the Abelianization of $\Gamma=\Delta(6n^2)$ is simply the group
generated by $A_1$.  Thus
\begin{equation}
\bar\Gamma=\mathbb Z_2,
\label{eq:z2}
\end{equation}
and is independent of any $n\mod3$ issues.

\subsection{The case $\Gamma=\Sigma(36)$}

The group $\Sigma(36)$ is generated by
\begin{equation}
V^4=A^3=B^3=1,
\end{equation}
along with
\begin{equation}
AV=VB,\qquad BV=VA^{-1}.
\end{equation}
Elements of $\Sigma(36)$ may be written as
\begin{equation}
g=V^vA^aB^b.
\end{equation}
Since the commutator subgroup is generated by
\begin{equation}
\{A,B\},
\end{equation}
the Abelianization of $\Gamma=\Sigma(36)$ is the group generated by $V$.
This gives
\begin{equation}
\bar\Gamma=\mathbb Z_4.
\end{equation}

\subsection{The case $\Gamma=\Sigma(36\times3)$}

The group $\Sigma(36\times3)$ is a central extension of $\Sigma(36)$.  It
is generated by
\begin{equation}
V^4=A^3=B^3=C^3=1,
\end{equation}
with $C$ a central extension, so that
\begin{equation}
AB=BAC,\qquad AV=VB,\qquad BV=VA^{-1}.
\end{equation}
Elements of $\Sigma(36\times3)$ may be written as
\begin{equation}
g=V^vA^aB^bC^c.
\end{equation}
Since $A$ and $B$ no longer commute, we ought to be a bit careful in
determining the generators of the commutator subgroup.  From the above
expressions, we see that the commutator subgroup is generated by
$\{A^{-1}B,B^{-1}A^{-1},C\}$.  Multiplying the first two generators in order
(and using $A^3=1$) gives $A$.  Since this is now part of the commutator
subgroup, it can be multiplied with $A^{-1}B$ to obtain $B$.  As a result,
the commutator subgroup is generated by
\begin{equation}
\{A,B,C\},
\end{equation}
so that it is in fact isomorphic to the Heisenberg group $\Delta(27)$.
Taking a quotient of $\{V,A,B,C\}$ by $\{A,B,C\}$ demonstrates that the
Abelianization of $\Gamma=\Sigma(36\times3)$ is the group generated by $V$.
Thus
\begin{equation}
\bar\Gamma=\mathbb Z_4.
\label{eq:z4}
\end{equation}
It is not particularly surprising that this is the same result as
the Abelianization of $\Sigma(36)$.

\section{The group theory of $\Delta(27)$ \label{sec:gtDelta27}}

Although the finite subgroups of $\mathrm{SU}(3)$ have been well studied
\cite{Bovier:1980gc,FFK,Blichfeldt}, we will try to make it accessible to
the careful reader by displaying the basic group theoretic properties used
in the above calculations.
Here, we consider the case where $\Gamma=\Delta(27)$.
This group is a single group in the series of groups $\Delta(3n^2)$
contained in $\mathrm{SU}(3)$.  We take the presentation of $\Delta(27)$
as follows:
\be
A^3=B^3=C^3=1, \quad AB=BAC, \quad AC=CA, \quad BC=CB.
\ee
This group has nine one-dimensional representations and two three-dimensional
representations.  These are given in Table~\ref{tbl:d27irrep}, where we
have chosen a convenient set of labels.

\begin{table}[t]
\begin{center}
\begin{tabular}{c||c|c|c}
Rep & $A$  &  $B$  &  $C$  \\
\hline \hline
$\bb{1}{0,0} $&$ 1   $&$ 1   $&$ 1  $\\
\hline
$\bb{1}{0,1}$ &$\gamma  $&$ 1   $&$ 1  $\\
\hline
$\bb{1}{0,2}$ &$\gamma^2  $&$ 1   $&$ 1  $\\
\hline
$\bb{1}{1,0}$ &$ 1   $&$\gamma  $&$ 1  $\\
\hline
$\bb{1}{1,1}$ &$ \gamma   $&$\gamma  $&$ 1  $\\
\hline
$\bb{1}{1,2}$ &$ \gamma^2   $&$\gamma  $&$ 1  $\\
\hline
$\bb{1}{2,0}$ &$1  $&$\gamma^2  $&$ 1  $\\
\hline
$\bb{1}{2,1}$ &$\gamma  $&$\gamma^2$&$ 1 $\\
\hline
$\bb{1}{2,2}$ &$\gamma^2  $&$\gamma^2$&$ 1 $\\
\hline
$\bb{3}{\;} $ &$ \begin{pmatrix}
               0 & 1 & 0 \\  0 & 0 & 1 \\
               1 & 0 & 0 \end{pmatrix}
$&$            \begin{pmatrix}
               1 & 0 & 0 \\
               0 &\gamma & 0 \\
               1 & 0 & \gamma^2 \end{pmatrix}
$&$            \begin{pmatrix}
               \gamma & 0 & 0 \\
               0 & \gamma & 0 \\
               0 & 0 & \gamma \end{pmatrix}   $ \\
\hline
$\bbbs{3} $ &$ \begin{pmatrix}
               0 & 1 & 0 \\  0 & 0 & 1 \\
               1 & 0 & 0 \end{pmatrix}
$&$            \begin{pmatrix}
               1 & 0 & 0 \\
               0 &\gamma^2 & 0 \\
               1 & 0 & \gamma \end{pmatrix}
$&$            \begin{pmatrix}
               \gamma^2 & 0 & 0 \\
               0 & \gamma^2 & 0 \\
               0 & 0 & \gamma^2 \end{pmatrix}   $
\end{tabular}
\end{center}
\caption{The eleven irreducible representations of $\Delta(27)$.  Here
$\gamma=e^{2\pi i/3}$ is a cube root of unity.}
\label{tbl:d27irrep}
\end{table}

Matching the eleven irreducible representations, the group $\Delta(27)$
also has eleven conjugacy classes.  This gives rise to the (partial)
character table
\begin{equation}
\begin{tabular}{c||c|c|c|c|c|c|c|c|c|c|c}
            & $1$  &  $C$  & $C^2$ & $A$   & $A^2$ & $B$   & $B^2$ & $AB$  & $A^2B$& $AB^2$& $A^2B^2$  \\
\hline \hline
$\bb{1}{0,0}$ &$ 1   $&$ 1   $&$ 1   $&$ 1   $&$ 1   $&$ 1   $&$ 1   $&$ 1   $&$ 1   $&$ 1  $&$ 1  $\\
\hline
$\bb{1}{0,1}$ &$ 1   $&$ 1   $&$ 1   $&$ \gamma   $&$ \gamma^2   $&$ 1   $&$ 1   $&$ \gamma   $&$\gamma^2   $&$\gamma $&$ \gamma^2 $ \\
\hline
$\bb{1}{1,0}$ &$ 1   $&$ 1   $&$ 1   $&$ 1   $&$ 1   $&$ \gamma   $&$ \gamma^2   $&$ \gamma  $&$ \gamma   $&$\gamma^2$&$ \gamma^2 $ \\
\hline
$\bb{1}{1,1}$ &$ 1   $&$ 1   $&$ 1   $&$ \gamma   $&$ \gamma^2   $&$ \gamma   $&$ \gamma^2   $&$ \gamma^2   $&$ 1   $&$ 1  $&$ \gamma $ \\
\hline
$\bb{1}{2,1}$ &$ 1   $&$ 1   $&$ 1   $&$ \gamma  $&$ \gamma^2   $&$ \gamma^2   $&$ \gamma   $&$ 1   $&$ \gamma   $&$ \gamma^2  $&$ 1 $ \\
\hline
$\bb{3}{\;}  $ &$ 3   $&$ 3\gamma   $&$ 3\gamma^2   $&$ 0   $&$ 0   $&$ 0   $&$ 0   $&$ 0   $&$ 0   $&$ 0 $&$ 0 $ \\
\end{tabular}
\end{equation}
The remaining characters are given by complex conjugation of the
above representations and characters.  Explicitly, we take
$\gamma \leftrightarrow \gamma^2$ in the characters, and map the labels
for the representations by $\bbs{3}\rightarrow {\bbbs{3}}$ and
$\bb{1}{i,j}\rightarrow\bb{1}{3-i,3-j}$, where these indices are taken mod~3.
Also, note that the conjugacy classes for the last eight columns above
contain three elements: the element given, as well as its product with $C$
and $C^2$.

The multiplication table for the above representations is given by
\begin{equation}
\begin{tabular}{c||c|c|c}
$\otimes$  & $\bb{1}{i,j}$&$\bb{3}{\;}$ & $\bIII$ \\
\hline \hline
$\bb{1}{k,l}^{\phantom{\frac{1}{1}}}$ & $\bb{1}{i+k,j+l}$ & $\bb{3}{\;}$
& $\bIII$ \\
\hline
$\bb{3}{\;}^{\phantom{\frac{1}{1}}}  $ & $\bb{3}{\;}$  &
$\bIII_{(\bar{1})}\oplus\bIII_{(\bar{2})}\oplus\bIII_{(\bar{3})}$ & $\oplus_{i,j}\bb{1}{i,j}$ \\
\hline
$\bIII^{\phantom{\frac{1}{1}}}  $ & $\bIII$  &
$\oplus_{i,j} \bb{1}{i,j} $ & $\bbs{3}_{({1})}\oplus\bbs{3}_{({2})}\oplus\bbs{3}_{({3})}$ \\
\end{tabular}
\end{equation}
For our purposes, we desire the exact form of the superpotential, which
may be obtained from the appropriate set of Clebsch-Gordan coefficients.
In particular, we will need the coefficients in the decomposition
$\mathbf{3}\otimes \bb{r}{i} \rightarrow \bb{r}{j}$.
We will label the functions that $\bb{1}{i,j}$ work on as $\phi_{i,j}$ and
the functions that $\mathbf3$ work
on as triplets $\Lambda_j$.  If multiple representations of the
same kind appear in a product or sum of representations, an index
in parentheses will appear on both the representation
as well as the functions associated with it.  The functions transforming
under a barred representation
will simply be labeled by putting a bar over the functions (although one
should not take this as complex conjugation).
In the following, we show the direct product and its resultant sum.  The
combinations of functions from the product that
the resultant works on in the way shown above is displayed after the colon.
For tensor products of the one-dimensional representations with $\mathbf3$,
we have
\bea
\bbs{3}\otimes \bb{1}{0,0}  &=& \bbs{3}:
    \begin{pmatrix} \Lambda_1\phi_{0,0} \\ \Lambda_2\phi_{0,0} \\ \Lambda_3\phi_{0,0}
    \end{pmatrix},  \nn \\
\bbs{3}\otimes \bb{1}{0,1}  &=& \bbs{3}:
    \begin{pmatrix} \Lambda_1\phi_{0,1} \\ \gamma\Lambda_2\phi_{0,1} \\
    \gamma^2\Lambda_3\phi_{0,1}
    \end{pmatrix},
  \quad
    \bbs{3}\otimes \bb{1}{0,2} = \bbs{3}:
    \begin{pmatrix} \Lambda_1 \phi_{0,2} \\
    \gamma^2\Lambda_2 \phi_{0,2} \\ \gamma\Lambda_3 \phi_{0,2}
    \end{pmatrix},  \nn \\
\bbs{3}\otimes \bb{1}{1,0}  &=& \bbs{3}:
    \begin{pmatrix} \Lambda_3\phi_{1,0} \\ \Lambda_1\phi_{1,0} \\
    \Lambda_2\phi_{1,0}
    \end{pmatrix},
   \quad\kern1em
    \bbs{3}\otimes \bb{1}{2,0} = \bbs{3}:
    \begin{pmatrix} \Lambda_2 \phi_{2,0} \\
    \Lambda_3 \phi_{2,0} \\ \Lambda_1 \phi_{2,0}
    \end{pmatrix},  \\
\bbs{3}\otimes \bb{1}{1,1} &=& \bbs{3}:
    \begin{pmatrix} \gamma^2\Lambda_3 \phi_{1,1} \\ \Lambda_1\phi_{1,1} \\
    \gamma\Lambda_2\phi_{1,1}
    \end{pmatrix},
   \quad
    \bbs{3}\otimes \bb{1}{2,2} = \bbs{3}:
    \begin{pmatrix} \Lambda_2\phi_{2,2} \\
    \gamma^2 \Lambda_3 \phi_{2,2} \\
    \gamma \Lambda_1 \phi_{2,2}
    \end{pmatrix},  \nn \\
\bbs{3}\otimes \bb{1}{2,1}  &=& \bbs{3}:
    \begin{pmatrix} \gamma\Lambda_2\phi_{2,1} \\ \gamma^2\Lambda_3\phi_{2,1} \\
    \Lambda_1\phi_{2,1}  \end{pmatrix},
   \quad
    \bbs{3}\otimes \bb{1}{1,2} = \bbs{3}:
    \begin{pmatrix} \Lambda_3\phi_{1,2} \\
    \gamma^2 \Lambda_1\phi_{1,2} \\
    \gamma \Lambda_2 \phi_{1,2}  \end{pmatrix}.  \nn
\eea
Similarly, one can work out how the one-dimensional representations act on
the $\bbbs{3}$ representation
\bea
\bbbs{3}\otimes \bb{1}{0,0}  &=& \bbbs{3}:
    \begin{pmatrix} \Lambda_1\phi_{0,0} \\ \Lambda_2\phi_{0,0} \\ \Lambda_3\phi_{0,0}
    \end{pmatrix},  \nn \\
\bbbs{3}\otimes \bb{1}{0,1}  &=& \bbbs{3}:
    \begin{pmatrix} \Lambda_1\phi_{0,1} \\ \gamma\Lambda_2\phi_{0,1} \\
    \gamma^2\Lambda_3\phi_{0,1}
    \end{pmatrix},
  \quad
    \bbbs{3}\otimes \bb{1}{0,2} = \bbbs{3}:
    \begin{pmatrix} \Lambda_1 \phi_{0,2} \\
    \gamma^2\Lambda_2 \phi_{0,2} \\ \gamma\Lambda_3 \phi_{0,2}
    \end{pmatrix},  \nn \\
\bbbs{3}\otimes \bb{1}{1,0}  &=& \bbbs{3}:
    \begin{pmatrix} \Lambda_2\phi_{1,0} \\ \Lambda_3\phi_{1,0} \\
    \Lambda_1\phi_{1,0}
    \end{pmatrix},
   \quad\kern1em
    \bbbs{3}\otimes \bb{1}{2,0} = \bbbs{3}:
    \begin{pmatrix} \Lambda_3 \phi_{2,0} \\
    \Lambda_1 \phi_{2,0} \\ \Lambda_2 \phi_{2,0}
    \end{pmatrix},  \\
\bbbs{3}\otimes \bb{1}{1,1} &=& \bbbs{3}:
    \begin{pmatrix} \gamma\Lambda_2 \phi_{1,1} \\ \gamma^2\Lambda_3\phi_{1,1} \\
    \Lambda_1\phi_{1,1}
    \end{pmatrix},
   \quad
    \bbbs{3}\otimes \bb{1}{2,2} = \bbbs{3}:
    \begin{pmatrix} \gamma \Lambda_3\phi_{2,2} \\
     \Lambda_1 \phi_{2,2} \\
    \gamma^2 \Lambda_2 \phi_{2,2}
    \end{pmatrix},  \nn \\
\bbbs{3}\otimes \bb{1}{2,1}  &=& \bbbs{3}:
    \begin{pmatrix} \gamma^2\Lambda_3\phi_{2,1} \\ \Lambda_1\phi_{2,1} \\
    \gamma \Lambda_2\phi_{2,1}  \end{pmatrix},
   \quad
    \bbbs{3}\otimes \bb{1}{1,2} = \bbbs{3}:
    \begin{pmatrix} \gamma^2 \Lambda_2\phi_{1,2} \\
    \gamma \Lambda_3\phi_{1,2} \\
    \Lambda_1 \phi_{1,2}  \end{pmatrix}.  \nn
\eea
Next, we display similarly the product of the $\bb{3}{\;}$ with itself:
\bea
\bb{3}{\;}^{(1)}\otimes \bb{3}{\;}^{(2)}&=&\bIII_{(\bar{1})}\oplus\bIII_{(\bar{2})}\oplus\bIII_{(\bar{3})}; \nn \\
\bIII_{(\bar{1})}&:&\begin{pmatrix} \Lambda_1^{(1)}\Lambda_1^{(2)} \\ \Lambda_2^{(1)}\Lambda_2^{(2)} \\ \Lambda_3^{(1)}\Lambda_3^{(2)} \end{pmatrix}, \nn \\
\bIII_{(\bar{2})}&:&\begin{pmatrix} \Lambda_2^{(1)}\Lambda_3^{(2)} \\ \Lambda_3^{(1)}\Lambda_1^{(2)} \\ \Lambda_1^{(1)}\Lambda_2^{(2)} \end{pmatrix}, \\
\bIII_{(\bar{3})}&:&\begin{pmatrix} \Lambda_3^{(1)}\Lambda_2^{(2)} \\ \Lambda_1^{(1)}\Lambda_3^{(2)} \\ \Lambda_2^{(1)}\Lambda_1^{(2)} \end{pmatrix}, \nn
\eea
and of $\bbs{3}$ with $\bIII$
\bea
\bbs{3}\otimes \bIII&=&\oplus_{i,j} \bb{1}{i,j}; \nn \\
\bb{1}{0,0}&:&\frac{1}{\sqrt{3}}
   \left(
   \Lambda_1 \bar{\Lambda}_{\bar{1}}+
   \Lambda_2 \bar{\Lambda}_{\bar{2}}+
   \Lambda_3 \bar{\Lambda}_{\bar{3}}\right),\nn \\
\bb{1}{0,1}&:&\frac{1}{\sqrt{3}}
   \left(
   \gamma \Lambda_1 \bar{\Lambda}_{\bar{1}}+
   \Lambda_2 \bar{\Lambda}_{\bar{2}}+
   \gamma^2 \Lambda_3 \bar{\Lambda}_{\bar{3}}\right),\nn \\
\bb{1}{0,2}&:&\frac{1}{\sqrt{3}}
   \left(
   \gamma^2 \Lambda_1 \bar{\Lambda}_{\bar{1}}+
   \Lambda_2 \bar{\Lambda}_{\bar{2}}+
   \gamma \Lambda_3 \bar{\Lambda}_{\bar{3}}\right),\nn \\
\bb{1}{1,0}&:&\frac{1}{\sqrt{3}}
   \left(
   \Lambda_1 \bar{\Lambda}_{\bar{3}}+
   \Lambda_2 \bar{\Lambda}_{\bar{1}}+
   \Lambda_3 \bar{\Lambda}_{\bar{2}}\right),\nn \\
\bb{1}{2,0}&:&\frac{1}{\sqrt{3}}
   \left(
   \Lambda_1 \bar{\Lambda}_{\bar{2}}+
   \Lambda_2 \bar{\Lambda}_{\bar{3}}+
   \Lambda_3 \bar{\Lambda}_{\bar{1}}\right),\nn \\
\bb{1}{1,1}&:&\frac{1}{\sqrt{3}}
   \left(
   \gamma\Lambda_1 \bar{\Lambda}_{\bar{3}}+
   \Lambda_2 \bar{\Lambda}_{\bar{1}}+
   \gamma^2\Lambda_3 \bar{\Lambda}_{\bar{2}}\right),\nn \\
\bb{1}{2,2}&:&\frac{1}{\sqrt{3}}
   \left(
   \gamma\Lambda_1 \bar{\Lambda}_{\bar{3}}+
   \gamma^2\Lambda_2 \bar{\Lambda}_{\bar{1}}+
   \gamma\Lambda_3 \bar{\Lambda}_{\bar{2}}\right),\nn \\
\bb{1}{2,1}&:&\frac{1}{\sqrt{3}}
   \left(
   \gamma\Lambda_1 \bar{\Lambda}_{\bar{2}}+
   \Lambda_2 \bar{\Lambda}_{\bar{3}}+
   \gamma^2\Lambda_3 \bar{\Lambda}_{\bar{1}}\right),\nn \\
\bb{1}{1,2}&:&\frac{1}{\sqrt{3}}
   \left(
   \Lambda_1 \bar{\Lambda}_{\bar{2}}+
   \gamma\Lambda_2 \bar{\Lambda}_{\bar{3}}+
   \gamma^2\Lambda_3 \bar{\Lambda}_{\bar{1}}\right).\nn
\eea
In addition to these, we always take the Clebsch-Gordan coefficients for
the product of two one-dimensional representations to be simply 1.  Also,
the Clebsch-Gordan coefficient associated with the product
$\bbs{r}\otimes \bbbs{r} \rightarrow \bb{1}{0,0}$
is always taken to be $(1/\sqrt{\dim{\bbs{r}}})\sum{f_i \bar{f}_i}$
in the usual way.

We take the expressions for the Clebsch-Gordan coefficients to be
commutative if two different representations are commutative, even though
the Kronecker product of two matrices is not.  Note that the above
Clebsch-Gordon coefficients are defined only up to a phase for each
representation in the sum.  This means that, although we have displayed
above that the phases of the $\bb{1}{0,1}$ and $\overline{(\bbs{1}_{0,1})}
=\bbs{1}_{0,2}$ representations appearing in $\bbs{3}\otimes \bIII$ are
related, they are in fact not related; we can independently rephase the two
representations. This corresponds directly to redefining the fields with
respect to their phase.  We have chosen these phases to make a certain
symmetry manifest in the resulting field theory.

We make one final note on this choice of Clebsch-Gordan coefficients.
While the Kronecker product of three matrices is associative, the above
choice for Clebsch-Gordan coefficients is not.  However, this does not
affect the physics. The above prescriptions define how to construct singlets
out of product representations.  Any method for arriving at these will be
physically equivalent, as the number of linearly independent singlet functions
is fixed for a given product. Therefore, the only possibility is that two
different conventions are related by some unitary transformation (taking
that both conventions involve unitary Clebsch-Gordan coefficients).


\end{document}

%% file: z3.pstex_t
\begin{picture}(0,0)%
\includegraphics{z3.pstex}%
\end{picture}%
\setlength{\unitlength}{3947sp}%
\begingroup\makeatletter\ifx\SetFigFont\undefined%
\gdef\SetFigFont#1#2#3#4#5{%
  \reset@font\fontsize{#1}{#2pt}%
  \fontfamily{#3}\fontseries{#4}\fontshape{#5}%
  \selectfont}%
\fi\endgroup%
\begin{picture}(2679,2261)(486,-1966)
\put(1887,-136){\makebox(0,0)[lb]{\smash{{\SetFigFont{8}{9.6}{\rmdefault}{\mddefault}{\updefault}{\color[rgb]{0,0,0}$\bbs{1}_0$}%
}}}}
\put(2782,-1580){\makebox(0,0)[lb]{\smash{{\SetFigFont{8}{9.6}{\rmdefault}{\mddefault}{\updefault}{\color[rgb]{0,0,0}$\bbs{1}_2$}%
}}}}
\put(1105,-1532){\makebox(0,0)[lb]{\smash{{\SetFigFont{8}{9.6}{\rmdefault}{\mddefault}{\updefault}{\color[rgb]{0,0,0}$\bbs{1}_1$}%
}}}}
\put(1943,130){\makebox(0,0)[lb]{\smash{{\SetFigFont{8}{9.6}{\rmdefault}{\mddefault}{\updefault}{\color[rgb]{0,0,0} }%
}}}}
\put(3112,-878){\makebox(0,0)[lb]{\smash{{\SetFigFont{8}{9.6}{\rmdefault}{\mddefault}{\updefault}{\color[rgb]{0,0,0} }%
}}}}
\put(1976,-1951){\makebox(0,0)[lb]{\smash{{\SetFigFont{8}{9.6}{\rmdefault}{\mddefault}{\updefault}{\color[rgb]{0,0,0} }%
}}}}
\put(501,-951){\makebox(0,0)[lb]{\smash{{\SetFigFont{8}{9.6}{\rmdefault}{\mddefault}{\updefault}{\color[rgb]{0,0,0} }%
}}}}
\end{picture}%

%% file: delta27.pstex_t
\begin{picture}(0,0)%
\includegraphics{delta27.pstex}%
\end{picture}%
\setlength{\unitlength}{3947sp}%
\begingroup\makeatletter\ifx\SetFigFont\undefined%
\gdef\SetFigFont#1#2#3#4#5{%
  \reset@font\fontsize{#1}{#2pt}%
  \fontfamily{#3}\fontseries{#4}\fontshape{#5}%
  \selectfont}%
\fi\endgroup%
\begin{picture}(3569,4131)(61,-3262)
\put(2529,-745){\makebox(0,0)[lb]{\smash{{\SetFigFont{8}{9.6}{\rmdefault}{\mddefault}{\updefault}{\color[rgb]{0,0,0}{\bf 1}$_{1,0}$}%
}}}}
\put(2569,567){\makebox(0,0)[lb]{\smash{{\SetFigFont{8}{9.6}{\rmdefault}{\mddefault}{\updefault}{\color[rgb]{0,0,0}{\bf 1}$_{0,0}$}%
}}}}
\put(2545,121){\makebox(0,0)[lb]{\smash{{\SetFigFont{8}{9.6}{\rmdefault}{\mddefault}{\updefault}{\color[rgb]{0,0,0}{\bf 1}$_{0,1}$}%
}}}}
\put(2529,-308){\makebox(0,0)[lb]{\smash{{\SetFigFont{8}{9.6}{\rmdefault}{\mddefault}{\updefault}{\color[rgb]{0,0,0}{\bf 1}$_{0,2}$}%
}}}}
\put(2529,-1174){\makebox(0,0)[lb]{\smash{{\SetFigFont{8}{9.6}{\rmdefault}{\mddefault}{\updefault}{\color[rgb]{0,0,0}{\bf 1}$_{1,1}$}%
}}}}
\put(2529,-1595){\makebox(0,0)[lb]{\smash{{\SetFigFont{8}{9.6}{\rmdefault}{\mddefault}{\updefault}{\color[rgb]{0,0,0}{\bf 1}$_{1,2}$}%
}}}}
\put(2536,-2008){\makebox(0,0)[lb]{\smash{{\SetFigFont{8}{9.6}{\rmdefault}{\mddefault}{\updefault}{\color[rgb]{0,0,0}{\bf 1}$_{2,0}$}%
}}}}
\put(2520,-2445){\makebox(0,0)[lb]{\smash{{\SetFigFont{8}{9.6}{\rmdefault}{\mddefault}{\updefault}{\color[rgb]{0,0,0}{\bf 1}$_{2,1}$}%
}}}}
\put(2504,-2882){\makebox(0,0)[lb]{\smash{{\SetFigFont{8}{9.6}{\rmdefault}{\mddefault}{\updefault}{\color[rgb]{0,0,0}{\bf 1}$_{2,2}$}%
}}}}
\put(270,-802){\makebox(0,0)[lb]{\smash{{\SetFigFont{8}{9.6}{\rmdefault}{\mddefault}{\updefault}{\color[rgb]{0,0,0}{\bf 3}}%
}}}}
\put(262,-1619){\makebox(0,0)[lb]{\smash{{\SetFigFont{8}{9.6}{\rmdefault}{\mddefault}{\updefault}{\color[rgb]{0,0,0}$\bar{\mbox{\bf 3}}$}%
}}}}
\put(1840,704){\makebox(0,0)[lb]{\smash{{\SetFigFont{8}{9.6}{\rmdefault}{\mddefault}{\updefault}{\color[rgb]{0,0,0} }%
}}}}
\put( 76,-1158){\makebox(0,0)[lb]{\smash{{\SetFigFont{8}{9.6}{\rmdefault}{\mddefault}{\updefault}{\color[rgb]{0,0,0}  }%
}}}}
\put(2270,-3247){\makebox(0,0)[lb]{\smash{{\SetFigFont{8}{9.6}{\rmdefault}{\mddefault}{\updefault}{\color[rgb]{0,0,0}    }%
}}}}
\put(3500,-1312){\makebox(0,0)[lb]{\smash{{\SetFigFont{8}{9.6}{\rmdefault}{\mddefault}{\updefault}{\color[rgb]{0,0,0}   }%
}}}}
\end{picture}%